\documentclass[twocolumn,showpacs,preprintnumbers,amsmath,amssymb,prb,aps]{revtex4}

\usepackage{epsfig}% Include figure files
\usepackage{graphicx}% Include figure files
\usepackage{dcolumn}% Align table columns on decimal point
\usepackage{bm}% bold math

\begin{document}

\title{Investigation of the spin state of Co in LaCoO$_3$ at room temperature}

\author{S.K. Pandey,$^{1,4}$ Ashwani Kumar,$^2$ S. Patil,$^1$
V.R.R. Medicherla,$^1$ R.S. Singh,$^1$ K. Maiti,$^1$ D.
Prabhakaran,$^3$ A.T. Boothroyd,$^3$ and A.V. Pimpale$^4$}

\affiliation{$^1$Department of Condensed Matter Physics and
Materials Science, Tata Institute of Fundamental Research, Homi
Bhabha Road, Colaba, Mumbai - 400 005, INDIA.\\
$^2$Department of Physics, Institute of Science and Laboratory
Education, IPS Academy, Indore 452 012, India.\\
$^3$Clarendon Laboratory, Department of Physics, University of
Oxford, Parks Road, Oxford OX1 3PU, UK.\\
$^4$UGC-DAE Consortium for Scientific Research, University Campus,
Khandwa Road, Indore 452 017, India.}

\date{\today}

\begin{abstract}

We investigate the spin state of LaCoO$_3$ using state-of-the-art
photoemission spectroscopy and {\em ab initio} band structure
calculations. The GGA+$U$ calculations provide a good description of
the ground state for the experimentally estimated value of electron
correlation strength, $U$. In addition to the correlation effect,
spin-orbit interaction is observed to play a significant role in the
case of intermediate spin and high spin configurations. The
comparison of the calculated Co 3$d$ and O 2$p$ partial density of
states with the experimental valence band spectra indicates that at
room temperature, Co has dominant intermediate spin state
configuration and that the high spin configuration may not be
significant at this temperature. The lineshape of the La 5$p$ and O
2$s$ core level spectra could be reproduced well within these {\em
ab initio} calculations.

\end{abstract}


\maketitle

\section{Introduction}

The evolution of the spin state of Co in LaCoO$_3$ with temperature
has drawn a great deal of attention during last 50
years.\cite{goodenough,heikes,raccah,bhide,veal,abbate,barman,lam,korotin,plakhty}
The ground state of this compound is believed to be nonmagnetic
(spin S = 0) insulator. It shows two magnetic transitions at about
100 K (a sharp transition) and at around 500 K
(broad).\cite{yamaguchi1} The second transition is also accompanied
by an insulator to metal transition. All these magnetic transitions
have been attributed to the temperature induced spin state
transition of Co$^{3+}$ ions. Initially it was believed that the
transition at 100~K occurs due to the change in the spin state of
Co$^{3+}$ ion from low spin (LS) state ($t_{2g}^6e_g^0 \Rightarrow
S$~=~0) to a mixed LS and high spin (HS) states ($t_{2g}^4e_g^2
\Rightarrow S$~=~2).\cite{goodenough,raccah} In order to achieve a
microscopic understanding of these transitions, photoemission
spectroscopy (PES) and $x$-ray absorption spectroscopy (XAS) have
been employed extensively, since these techniques help to probe the
electronic structure directly.\cite{abbate,barman,lam} The
experimental results were simulated using configuration interaction
(CI) calculations for the cluster of CoO$_6$ octahedron. All these
investigations inferred the presence of varying mixtures of LS and
HS states above 100 K.

Subsequently, a detailed study\cite{korotin} based on LDA+$U$ (LDA =
local density approximation and $U$ = electron-electron Coulomb
repulsion strength) calculations, attributed the 100~K transitions
to LS to orbital ordered intermediate spin (IS) state
($t_{2g}^5e_g^1 \Rightarrow S$~=~1) and the one at 500~K to orbital
disordered IS state as the energy of the latter state was found to
be higher than the former and lower than HS state. Since then,
numerous experimental and theoretical works have been carried out,
which attributed the first transition to LS to IS
state.\cite{kobayashi,zobel,radaelli,maris,nekrasov,ishikawa,phelan}
This has also been demonstrated by PES and XAS studies,\cite{saitoh}
where Co 2$p$ core level, valence band and O $K$-edge XAS spectra of
LaCoO$_3$ revealed 70\% LS and 30\% IS states in the temperature
range 100~K to 300~K. No contribution of HS state was observed up to
300~K.

Recently, based on GGA+$U$ (GGA = generalized gradient
approximation) calculations, Kn$\acute{i}$zek {\it et al}. showed
that the mixed LS-HS state (1:1) is the first excited state and thus
attributed the first magnetic transition from LS to LS-HS
state.\cite{knizek} This is in sharp contrast to the belief of IS
contributions in the intermediate temperature range. In addition,
experimental results based on XAS and magnetic circular
dichroism\cite{haverkort} and inelastic neutron
scattering\cite{podlesnyak} also interpreted in terms of mixed LS-HS
state in the temperature range up to 700~K considering cluster
approximations. It is, thus, evident that the spin state of Co at
different temperatures is still controversial.

In this work, we investigate the spin state of Co in LaCoO$_3$ at
room temperature using state-of-the-art photoemission spectroscopy
and {\em ab initio} band structure calculations. We observe that
spin-orbit coupling (SOC) plays an important role in determining the
electronic structure in this system. The comparison of the high
resolution spectra and the calculated results suggest that the
electronic structure of LaCoO$_3$ at room temperature has large IS
contributions. The {\em ab initio} calculations also provide a good
representation of the experimental La 5$p$ and O 2$s$ core level
spectra.

\section{Experimental and computational details}

Single crystal of LaCoO$_3$ was grown and characterized as described
elsewhere.\cite{prabhakaran} The photoemission spectra of the
valence band and shallow core levels were recorded at room
temperature (RT) using a spectrometer equipped with monochromatic Al
$K\alpha$ (1486.6 eV) and He {\scriptsize I} (21.2 eV) sources, and
a Gammadata Scienta analyzer, SES2002. The base pressure during the
measurements was about 4$\times$10$^{-11}$ Torr. The energy
resolution for the $x$-ray photoemission was set to 0.3 eV for the
valence band and 0.6 eV for the core level spectra. The resolution
for the He {\scriptsize I} spectrum was fixed to 4 meV. The sample
was cleaned {\em in situ} by scraping the sample surface using a
diamond file. The cleanliness of the sample was ascertained by
tracking the sharpness of O 1$s$ peak and absence of C 1$s$ peak.
The Fermi level was aligned by recording the valence band spectrum
of an Ag foil mounted on the same sample holder.

The LDA+$U$ and GGA+$U$ ($U$ represents the electron correlation
strength among Co 3$d$ electrons), spin-polarized density of states
(DOS) calculations were carried out using LMTART 6.61 (Ref. 24). For
calculating charge density, full-potential linearized Muffin-Tin
orbital (LMTO) method working in plane wave representation was
employed. In the calculation, we have used the Muffin-Tin radii of
3.515, 2.005, and 1.64 a.u. for La, Co and O, respectively. The
charge density and effective potential were expanded in spherical
harmonics up to $l$ = 6 inside the sphere and in a Fourier series in
the interstitial region. The initial basis set included 6$s$, 5$p$,
5$d$, and 4$f$ valence, and 5$s$ semicore orbitals of La; 4$s$,
4$p$, and 3$d$ valence, and 3$p$ semicore orbitals of Co, and 2$s$
and 2$p$ orbitals of O. The exchange correlation functional of the
density functional theory was taken after H.S. Vosko {\em et al}.
(Ref. 25) and GGA calculations are performed following J.P. Perdew
{\em et al.} (Ref. 26). The calculations were performed by taking
LS, IS and HS configurations, which correspond to ($t_{2g\uparrow}^3
t_{2g\downarrow}^3$), ($t_{2g\uparrow}^3 t_{2g\downarrow}^2
e_{g\uparrow}^1$) and ($t_{2g\uparrow}^3 t_{2g\downarrow}^1
e_{g\uparrow}^2$) electronic configurations, respectively as initial
input. Self-consistency was achieved by demanding the convergence of
the total energy to be smaller than 10$^{-4}$ Ryd/cell. Final
orbital occupancies for Co $t_{2g}$ and $e_g$ states were obtained
from self-consistent GGA+$U$ calculations for different initial
state configurations. (8, 8, 6) divisions of the Brillouin zone
along three directions for the tetrahedron integration were used to
calculate the DOS.

\section{Results and discussions}

\begin{figure}
\vspace{-2ex}
%\begin{center}
\includegraphics[angle=0,width=0.4\textwidth]{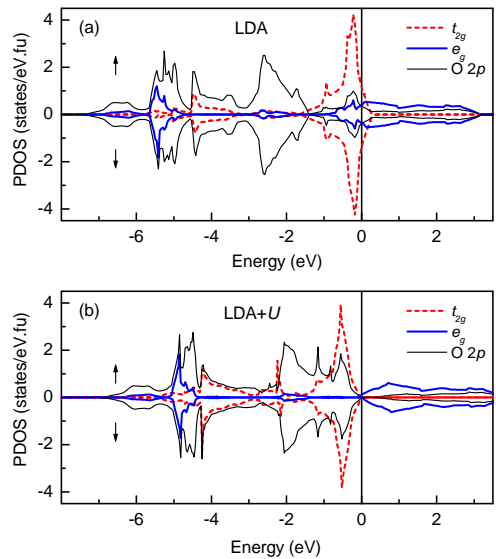}
\vspace{-10ex}
%\end{center}
\caption{(color online) O 2$p$ (thin solid line) and Co 3$d$ partial
density of states having $t_{2g}$ (dashed line) and $e_g$ (thick
solid line) symmetries from (a) LDA and (b) LDA+$U$ calculations.}
 \vspace{-2ex}
\end{figure}

It is well known that although the electron correlation effects are
underestimated in {\em ab initio} band structure calculations within
LDA, these results are often found to explain most of the features
in PES and XAS results in transition metal oxide
systems.\cite{ddPRL,kbmPRB1,kbmPRBR,pandey1,pandey2} However, one
needs to consider the correlation effects to capture the details of
the spectra and the ground state properties of the system. We have
calculated various contributions in the DOS of LaCoO$_3$ using both
LDA and LDA+$U$ methods. In order to minimize uncertainty in the
calculated results, we have fixed the value of $U$ to the
experimentally estimated values of 3.5~eV.\cite{ashish}

In Fig.~1(a), we show the calculated LDA results for both the up and
down spin states. There are several features observed in different
energy positions. The features between -1.5 eV to -3 eV energies are
dominated by the O 2$p$ states with negligible contribution from
other electronic states. These features are identified as O 2$p$
non-bonding states. The features at lower energies have dominant
contributions from O 2$p$ electronic states with small but finite
intensities from the Co 3$d$ states having $t_{2g}$ and $e_g$
symmetries. Thus, the features between -3 to -4.5 eV can be
attributed to bonding states with $t_{2g}$ symmetry and those below
-4.5 eV are the bonding states with $e_g$ symmetry. The antibonding
features appear above -1.5 eV and have predominantly Co 3$d$
character. Co 3$d$ bands with $t_{2g}$ symmetry appear between -1.5
to 0.5 eV and the $e_g$ bands between -1 to 3 eV. It is evident that
the density of states at the Fermi level is large suggesting a
metallic phase in the ground state in contrast to the insulating
phase observed in various experiments.

The effect of electron correlation in the density of states is
manifested in the LDA+$U$ results as shown in Fig. 1(b). Although
electron correlation among O 2$p$ electrons is not considered in the
calculations, the energy distribution of the O 2$p$ partial density
of states (PDOS) appears to be somewhat different from the LDA
results. The O 2$p$ contribution in the antibonding region appears
to enhance in the LDA+$U$ results compared to the LDA results
suggesting a spectral weight transfer from the bands having bonding
character. The non-bonding O 2$p$ contributions also shift toward
the Fermi level. The changes in the Co 3$d$ bands are most
significant as expected. The bonding and antibonding $e_g$ bands
shift towards higher energies. The bonding $t_{2g}$ band appears
almost at the same energies along with a significant increase in the
spectral weight. Subsequently, the antibonding $t_{2g}$ band having
primarily Co 3$d$ character becomes narrower along with a reduction
in spectral weight. It is, thus, clear that consideration of $U$ in
the calculations leads to a decrease in Co 3$d$ character and an
increase in O 2$p$ character of the bands in the vicinity of the
Fermi level. Although the overlap between the $t_{2g}$ and $e_g$
band is minimized for these parameters, no hard gap is observed
characterizing the systems to be metallic.

\begin{figure}
\vspace{-2ex}
%\begin{center}
\includegraphics[angle=0,width=0.4\textwidth]{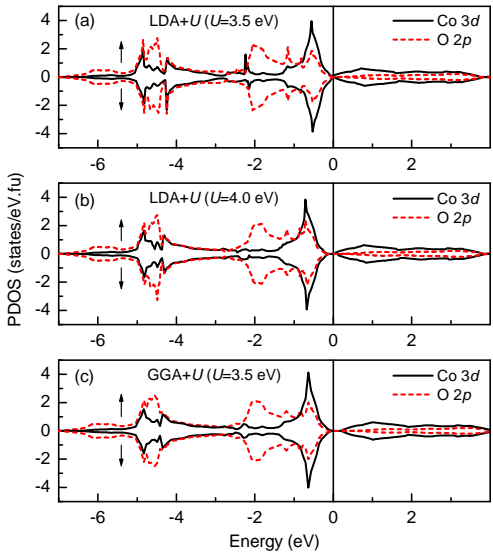}
\vspace{-10ex}
%\end{center}
\caption{(color online)  Co 3$d$ and O 2$p$ partial density of
states calculated using (a) LDA+$U$ ($U$ = 3.5 eV), (b) LDA+$U$ ($U$
= 4 eV) and (c) GGA+U ($U$ = 3.5 eV) methods.}
 \vspace{-2ex}
\end{figure}

In order to investigate the value of $U$ that creates a gap at the
Fermi level, we compare the Co 3$d$ and O 2$p$ PDOS for $U$ = 3.5 eV
and 4 eV in Fig. 2(a) and 2(b), respectively. It is evident that, an
increase in $U$ to 4 eV generates a band gap of about 0.2 eV. The
shape and energy distribution of the DOS is almost the same in both
the cases. The GGA is known to provide a better description of the
exchange-correlation functional as it also considers first order
correction to the spatial distribution of the electronic charge
density used in the LDA calculations. In GGA, total energy of the
system improves\cite{perdew} and it is expected that the energy
position of different bands would also improve. The calculated PDOS
of Co 3$d$ and O 2$p$ corresponding to GGA+$U$ calculations are
shown in Fig.~2(c). In contrast to the metallic phase observed in
LDA results, a value of $U$ = 3.5 eV creates a band gap of about
0.22 eV at the Fermi level in the results corresponding to GGA
calculations, which is very close to that observed
experimentally.\cite{arima,iguchi} However, the shape and positions
of Co 3$d$ and O 2$p$ bands is very similar in the LDA+$U$ and
GGA+$U$ results. Thus, we used GGA+$U$ method in the rest part of
our study to discuss the experimental results.

\begin{figure}
\vspace{-2ex}
%\begin{center}
\includegraphics[angle=0,width=0.4\textwidth]{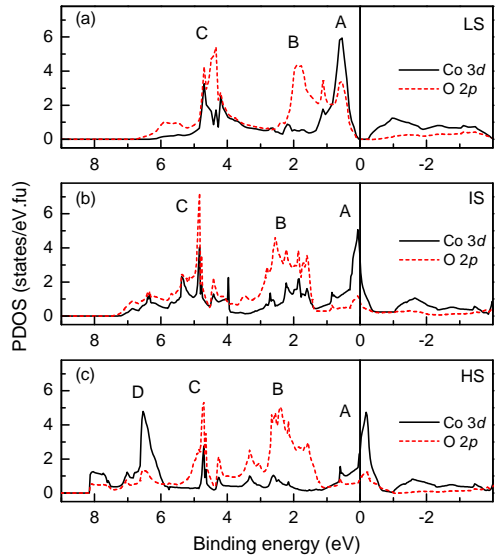}
\vspace{-10ex}
%\end{center}
\caption{(color online) Calculated Co 3$d$ and O 2$p$ partial
density of states corresponding to low spin (LS), intermediate spin
(IS) and high spin (HS) configurations using GGA+$U$ method.}
 \vspace{-2ex}
\end{figure}

We now turn to the question of the influence of various spin states
in the electronic structure of LaCoO$_3$. The PDOS obtained for
different spin state configurations are shown in Fig.~3 for the same
values of $U$ (= 3.5 eV). It is evident that although the LS
configuration leads to insulating ground state, the electronic
structure corresponding to IS and HS configuration converges to
metallic phase. This is a well know fact for this system and one
needs to consider additional parameters such as orbital ordering to
achieve insulating phase.\cite{korotin} However, the energy
distribution of the DOS over the large binding energy scale remains
very similar. There are three distinctly separable features A, B and
C in the DOS in Fig.~3. In the case of IS configuration, the
$t_{2g}$ band becomes partially filled. This effect enhances further
in the case of HS configuration. The total intensity of the feature
A gradually decreases with the change in spin state configuration
from LS $\rightarrow$ IS $\rightarrow$ HS. In addition, the Co 3$d$
character of the feature A becomes highest in the IS configuration
(80.3\%) while it is about 63.8\% in the LS state and 64.8\% in the
HS state. The feature B representing the non-bonding O 2$p$ states
appears at slightly higher binding energies in the IS and HS states
compared to that in the LS state. The change in the feature C is
again substantial. The O 2$p$ and Co 3$d$ contributions are almost
similar in LS and IS cases. In the case of HS states, the feature C
has dominant O 2$p$ character and a new feature D having dominant Co
3$d$ character appears beyond 6 eV binding energies.

\begin{figure}
\vspace{-2ex}
%\begin{center}
\includegraphics[angle=0,width=0.4\textwidth]{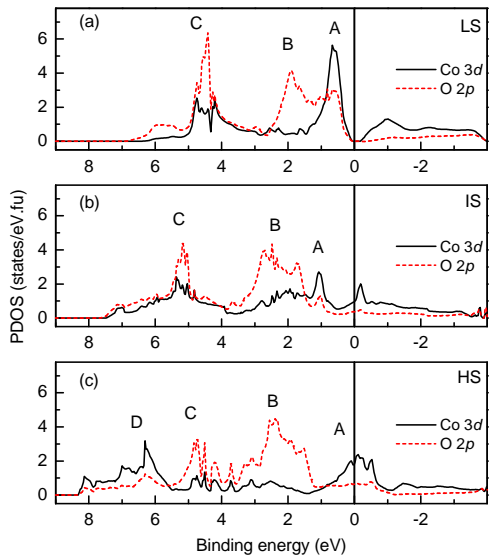}
\vspace{-8ex}
%\end{center}
\caption{(color online) Co 3$d$ and O 2$p$ partial density of states
corresponding to low spin (LS), intermediate spin (IS) and high spin
(HS) configurations calculated including spin-orbit interactions.
All these calculations are performed using GGA+$U$ method.}
 \vspace{-2ex}
\end{figure}

All the above calculations are done without considering the
spin-orbit coupling (SOC) in any of the electronic states. Recently,
there is a growing realization that SOC plays crucial role in
determining the electronic structure in various transition metal
oxides. The importance of SOC has been demonstrated in the systems
possessing rare-earth 4$f$ electrons.\cite{hotta} In Ca$_2$RuO$_4$,
it was shown that SOC plays an important role in the changeover of
the spin and orbital anisotropy as a function of
temperature.\cite{mizokawa} Very recent theoretical studies of
Ca$_3$CoRhO$_6$ compound have shown that the insulating ground state
can be achieved by considering SOC in the calculations.\cite{wu} In
order to investigate the effect in the present case, we have
calculated the PDOS in all the cases including SOC. The results are
shown in Fig.~4. It is clear that SOC has no significant influence
in the case of LS configuration. This is presumably due to the fact
that all the energy bands are completely filled. The effect is most
pronounced in Fig.~4(b) corresponding to IS state exhibiting a
splitting of the sharp feature A.

\begin{table}
%\vspace{6ex}
\caption{Occupancies of different Co 3$d$ orbitals obtained from
GGA+$U$ calculations for low spin (LS), intermediate spin (IS) and
high spin (HS) configurations of Co$^{3+}$ in LaCoO$_3$. Numbers
written in normal and bold face correspond to calculation without
spin-orbit coupling and with spin-orbit coupling, respectively. }
\vspace{2ex}
\begin{ruledtabular}
\begin{tabular}{|c|c|c|c|c|c|}
Initial spin state & $t_{2g\uparrow}$ & $e_{g\uparrow}$ &
$t_{2g\downarrow}$ & $e_{g\downarrow}$ & Total \\ \hline

LS & 2.77 & 0.58 & 2.77 & 0.58 & 6.7 \\

   & {\bf 2.77} & {\bf 0.59} & {\bf 2.77} & {\bf 0.58} & {\bf 6.71}
   \\ \hline

IS & 2.79 & 1.5 & 1.95 & 0.43 & 6.67 \\

   & {\bf 2.8} & {\bf 1.36} & {\bf 2.03} & {\bf 0.45} & {\bf 6.64}
   \\ \hline

HS & 2.82 & 1.9 & 1.32 & 0.42 & 6.46 \\

   & {\bf 2.83} & {\bf 1.9} & {\bf 1.35} & {\bf 0.41} & {\bf 6.49}
   \\

\end{tabular}
\end{ruledtabular}
\end{table}

The effect of spin-orbit coupling in the occupancies of
$t_{2g\uparrow}$ , $e_{g\uparrow}$ , $t_{2g\downarrow}$, and
$e_{g\downarrow}$ are given in Table 1. Numbers written in normal
and bold face correspond to without SOC and with SOC, respectively.
The total electronic occupancies of Co 3$d$ bands corresponding to
LS, IS and HS states are about 6.7, 6.65, and 6.47, respectively.
These values are closer to those obtained from full-multiplet
configuration interaction calculations carried out by Saitoh {\em et
al}. (Ref. 19) and somewhat smaller from those obtained by Korotin
{\em et al.} (Ref. 9). It is evident that the total occupancies as
well as partial occupancies of $t_{2g}$ and $e_g$ orbitals
corresponding to LS and HS states are not affected significantly by
SOC. However, in IS state, the occupancies of $e_{g\uparrow}$
orbitals decreases by 0.14 and that of $t_{2g\downarrow}$ orbital
increases by 0.08 due to the spin-orbit coupling keeping the total
occupancy almost unchanged.

\begin{figure}
\vspace{-2ex}
%\begin{center}
\includegraphics[angle=0,width=0.4\textwidth]{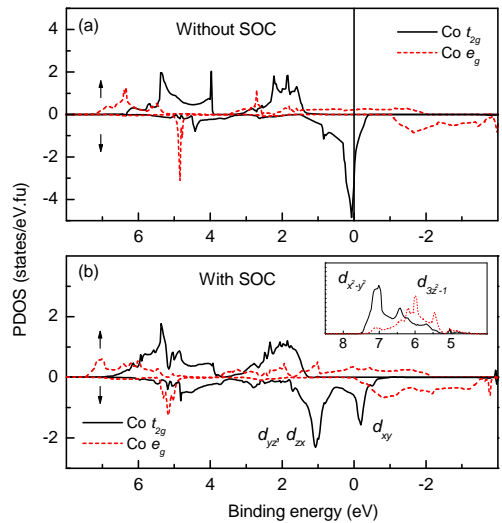}
\vspace{-12ex}
%\end{center}
\caption{(color online) Spin polarized partial density of states
corresponding to Co $t_{2g}$ and $e_g$ bands calculated (a) without
spin-orbit coupling and (b) with spin-orbit coupling. The inset
shows $d_{x^2-y^2}$ and $d_{z^2}$ contributions in the
$e_{g\uparrow}$ band.}
 \vspace{-2ex}
\end{figure}

The details of the effect of SOC in the IS state are shown in
Fig.~5, where we show the up and down spin PDOS of Co 3$d$ bands
separately. In Fig. 5(a), the feature in the vicinity of the Fermi
level arises due to the down spin $t_{2g}$ states. This feature
splits into two distinctly separated features (the energy separation
is about 1.2 eV). The feature at about 1 eV binding energy
represents the doubly degenerate $d_{xz}$ and $d_{yz}$ bands and the
non-degenerate $d_{xy}$ band appears above the Fermi level. The
doubly degenerate bonding $e_{g\uparrow}$ band also splits by about
1 eV into $d_{x^2-y^2}$ and $d_{z^2}$ bands as shown in the inset of
Fig.~5(b). It is thus clear that the degeneracy in the $t_{2g}$ and
$e_g$ bands are partially lifted due to spin-orbit interactions and
the effect is most pronounced in the IS configuration. Since, the
$d$ bands are not completely filled, the splitting of the $t_{2g}$
and $e_g$ bands will influence the occupancy of these bands. Here,
the SOC induced splitting leads to lowering in energy of $d_{yz}$ ,
$d_{zx}$ and $d_{x^2-y^2}$ bands and enhances the energy of $d_{xy}$
and $d_{z^2}$ bands. Thus, the occupancies of $d_{yz}$ , $d_{zx}$
and $d_{x^2-y^2}$ are expected to increase and that of $d_{xy}$  and
$d_{z^2}$ to decrease. This is clearly manifested in the partial
occupancies of each Co 3$d$ orbitals given in Table 2, where the
occupancies of $d_{yz\downarrow}$ , $d_{zx\downarrow}$ and
$d_{(x^2-y^2)\uparrow}$ increase by 0.27, 0.27 and 0.11,
respectively, and that of $d_{xy\downarrow}$ and $d_{z^2\uparrow}$
decrease by 0.46 and 0.15, respectively.

\begin{table}
%\vspace{12ex}
\caption{Occupancies of 3$d$ orbitals in intermediate spin
configuration of Co$^{3+}$ in LaCoO$_3$. Numbers written in normal
and bold face are corresponding to calculation without spin-orbit
coupling and with spin-orbit coupling, respectively. }
 \vspace{2ex}
\begin{ruledtabular}
\begin{tabular}{|c|c|c|c|c|c|}
Spin & $d_{yz}$ & $d_{xz}$ & $d_{xy}$ & $d_{x^2-y^2}$ & $d_{z^2}$
\\ \hline

Up & 0.93 & 0.93 & 0.93 & 0.75 & 0.75 \\

   & {\bf 0.93} & {\bf 0.93} & {\bf 0.94} & {\bf 0.86} & {\bf 0.50}
   \\ \hline

Down & 0.65 & 0.65 & 0.65 & 0.21 & 0.21 \\

   & {\bf 0.92} & {\bf 0.92} & {\bf 0.19} & {\bf 0.22} & {\bf 0.23}
   \\

\end{tabular}
\end{ruledtabular}
\end{table}

In order to compare the calculated results with the experimental
spectra, we show the background subtracted experimental valence band
spectra at RT in Fig.~6. The spectra show three distinct features at
about 1.0, 2.9 and 5.3 eV binding energies denoted by A, B and C,
respectively. The relative intensity of various features are
significantly different in the spectra using Al $K\alpha$ and He
{\scriptsize I} sources. While the feature A is most intense in the
Al $K\alpha$ spectrum, it becomes weakest in intensity in the He
{\scriptsize I} spectrum. Subsequently, the intensities of the
features B and C become dominant in the He {\scriptsize I} spectrum.
It is well known that the photoemission cross section is very
sensitive to the excitation energies.\cite{yeh} While the cross
section of O 2$p$ states is larger than that of Co 3$d$ states in
the He {\scriptsize I} spectrum, the relative cross section becomes
opposite in the Al $K\alpha$ energies making the cross section for
Co 3$d$ states significantly larger than that of O 2$p$ states.
Thus, the changes in intensity of various features due to the change
in excitation energy suggests that the feature A is predominantly
contributed by Co 3$d$ states and the features B and C have large O
2$p$ character. No significant feature is observed between 6 to 8 eV
binding energies. This clearly suggests that the experimental
results are very similar to the results obtained for IS
configuration shown in Fig.~4(b).

\begin{figure}
\vspace{-2ex}
%\begin{center}
\includegraphics[angle=0,width=0.4\textwidth]{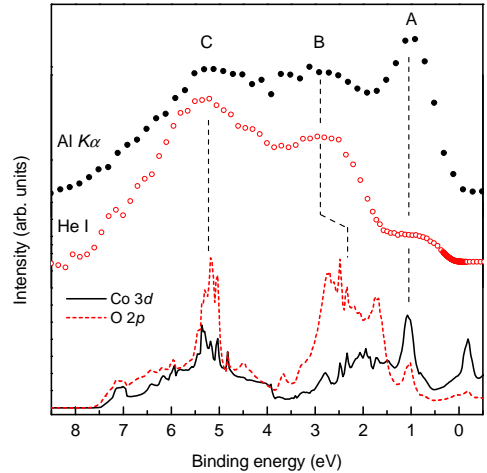}
\vspace{-16ex}
%\end{center}
\caption{(color online) Experimental valence band spectra collected
at room temperature using Al $K\alpha$ and He {\scriptsize I}
radiations. The lines denote the calculated Co 3$d$ and O 2$p$
partial density of states corresponding to intermediate spin
configuration of Co including spin-orbit coupling.}
 \vspace{-2ex}
\end{figure}

In order to bring out clarity to this comparison, we plot the O 2$p$
and Co 3$d$ PDOS corresponding to IS configuration in Fig.~6. The
peak position of the feature B is somewhat higher (0.5 eV) in the
experimental spectra compared to the calculated results. Such small
shift in the completely filled non-bonding O 2$p$ bands has often
been observed in the LDA results due to the underestimation of the
electron correlation among the O 2$p$ electrons.\cite{ddPRL} The
energy positions of the features A and C in the experimental
spectra, and the large Co 3$d$ character of the feature A and O 2$p$
character of feature C is revealed remarkably in the calculated
results. Even the small shoulder in the binding energy range 5.5 -
7.5 eV with relatively larger O 2$p$ character is reproduced in the
calculated results. These results clearly suggest that the
electronic structure of LaCoO$_3$ at room temperature corresponds
primarily to the intermediate spin state configuration of Co and the
contribution from HS state configuration may not be so significant.
These results appear to be different from those in the recent
studies based on cluster approximations.\cite{haverkort} Such
difference may not be surprising as in the cluster calculations, the
Co 3$d$ and O 2$p$ bands are approximated to atomic energy levels.

\begin{figure}
\vspace{-2ex}
%\begin{center}
\includegraphics[angle=0,width=0.45\textwidth]{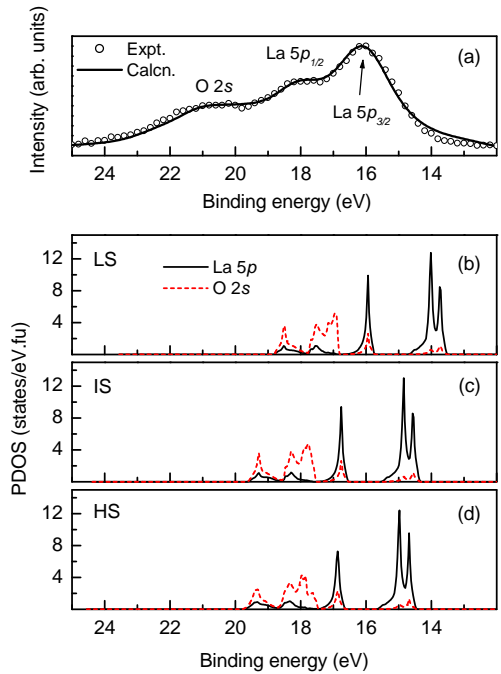}
\vspace{-8ex}
%\end{center}
\caption{(color online) (a) Background subtracted La 5$p$ and O 2$s$
core level spectra (symbols) collected at room temperature. The
solid line represent the spectrum simulated using PDOS corresponding
to IS state configuration. La 5$p$ (solid line) and O 2$s$ (dashed
line) PDOS calculated using (b) LS, (c) IS and (d) HS configurations
of Co.}
 \vspace{-2ex}
\end{figure}

Finally, we investigate the applicability of {\em ab initio} band
structure calculations in understanding the shallow core level
spectrum. Such study will be quite interesting as core-hole
potential is known to influence the final states of the core level
photoemission. The integral background subtracted experimental
spectrum along with the calculated one is plotted in Fig.~7(a).
Three distinct features visible in the spectrum. The features at 16
and 18 eV binding energies correspond to spin-orbit split La 5$p$
states as marked in the figure. The feature around 21 eV binding
energy represent the signature of photoemission from O 2$s$ levels.

We also show the La 5$p$ and O 2$s$ PDOS corresponding to LS, IS and
HS configurations in different panels of Fig.~7. All the three
features are present in the calculated results corresponding to LS,
IS and HS states. The shape of the bands in each region remains
almost the same in each spin state. The spin-orbit splitting of the
La 5$p$ contributions in the PDOS is about 2 eV as also observed in
the experimental spectra. However, the peak position of the bands
corresponding to LS state appears at binding energies lower by about
1 eV in comparison to energies corresponding to IS and HS states,
which are closer to the experimental spectra.

In order to simulate the experimental spectrum, we have shifted the
calculated O 2$s$ and La 5$p$ PDOS obtained from IS configuration by
3.1 eV and 1.2 eV, respectively and then multiplied by photoemission
cross sections of the corresponding states.\cite{yeh} To account for
the lifetime broadening and resolution broadening, O 2$s$ and La
5$p$ PDOS are convoluted with a suitable Lorentzian and a Gaussian,
respectively. The simulated spectrum is shown by solid line
overlapped on the experimental spectrum in Fig.~7(a). The
reproduction of the experimental spectrum is remarkable. This
suggests that one can indeed employ band structure calculations to
determine the core level spectrum. It is to note here that these
calculations cannot create features appearing due to different
screening effects expected in the final state of photoemission.
Almost perfect reproduction of the experimental spectrum in the
present case indicates that final state effects may not be
significant in the case of La 5$p$ photoemission.

\section{Conclusions}

In summary, we have investigated the electronic structure of
LaCoO$_3$ at room temperature using various forms of {\em ab initio}
calculations and high resolution photoemission spectroscopy on high
quality single crystal. We observe that GGA+$U$ calculations provide
a good description of the ground state. Spin-orbit coupling appears
to play a significant role in the case of IS and HS states of Co,
which is most pronounced in the case of IS state of Co. The
calculated Co 3$d$ and O 2$p$ partial density of states
corresponding to intermediate spin state of Co provide the best
description of the experimental valence band spectra at room
temperature. This suggests that the spin state of Co at room
temperature presumably has dominant intermediate spin configuration
and that the contribution from high spin state is not so significant
at this temperature. The applicability of the {\em ab initio} band
structure calculations in understanding the shallow core level
spectrum is also studied. The calculated exchange splitting of La
5$p$ states is found to be about 2 eV, which is identical to the
experimentally observed splitting. The lineshape of the calculated
spectrum provides a good description of the experimental spectrum.
This suggests that final state effects may not be significant in the
case of La 5$p$ photoemission in LaCoO$_3$.

\section{Acknowledgements}

The authors would like to thank Prof. A. V. Narlikar for continued
support. SP is thankful to CSIR, India, for financial support.

\end{document}